\documentclass[conference]{IEEEtran}
\IEEEoverridecommandlockouts

\usepackage{cite}
\usepackage{url}
\usepackage{hyperref}
\usepackage{amsmath,amssymb,amsfonts}
\newtheorem{definition}{Definition}
\usepackage{graphicx}
\usepackage{textcomp}
\usepackage{xcolor}
\usepackage{multirow}
\usepackage[linesnumbered,ruled,vlined]{algorithm2e}
\usepackage{algpseudocode}
\let\oldnl\nl
\newcommand{\nonl}{\renewcommand{\nl}{\let\nl\oldnl}}
\def\BibTeX{{\rm B\kern-.05em{\sc i\kern-.025em b}\kern-.08em
    T\kern-.1667em\lower.7ex\hbox{E}\kern-.125emX}}
\begin{document}

\title{Learning Red Agent Policy from Observations for
Neurosymbolic Autonomous Cyber Agents
}

\author{\IEEEauthorblockN{Ankita Samaddar, Sandeep Neema, Daniel Balasubramanian, Xenofon Koutsoukos}
\IEEEauthorblockA{\textit{Departmaent of Computer Science} \\
\textit{Vanderbilt University}\\
Nashville, TN \\
\{ankita.samaddar, sandeep.neema, daniel.a.balasubramanian, xenofon.koutsoukos\}@vanderbilt.edu}
}

\maketitle

\begin{abstract}
With sophisticated cyber-attacks becoming increasingly prevalent, modern networks require intelligent autonomous cyber-defense agents trained via Reinforcement Learning (RL). These agents employ neurosymbolic approaches such as behavior trees with learning-enabled components (LECs) to learn, reason, adapt, and implement security rules while maintaining critical operations. However, these autonomous networks are partially observable systems, i.e., the cyber-attacker's (red agent's) actions are not observable, making it difficult for the defender to predict red actions, learn red policies, or assess the attacker's intrusion levels. To address this, we propose a \textit{Policy Learning Technique} using imitation learning to learn policies for partially observable
RL agents with discrete states and discrete actions. We apply this technique in an autonomous cyber environment to predict red agent's actions from network observations and defender actions. Integrated with a neurosymbolic cyber-defense agent, our method effectively handles different red policies and achieves high prediction accuracy across diverse simulated scenarios.
\end{abstract}

\section{Introduction}\label{sec:intro}
\noindent
Modern cyber applications take the advantage of autonomous networks to automate workflows by deploying agents that independently learn and enforce security rules to defend against cyber-attacks~\cite{kott2023autonomousintelligentcyberdefenseagent}. Achieving this defense necessitates continuous system monitoring, early breach detection, and timely selection of appropriate countermeasures to contain the breaches, while ensuring uninterrupted operational workflows. Conventional security standards, however, often prove inadequate against sophisticated adversarial strategies. To address this limitation, a hybrid approach is needed that combines security standards with learning enabled components (LECs). These LECs are often reinforcement learning (RL) based function approximators enabling the cyber-defense agents to adapt dynamically and take optimal actions against evolving threats.

In autonomous cyber environments, the network evolves from one state to another under the combined influence of red (attacker) and blue (defender) actions. However, the environment is partially observable, \emph{i.e.}, the defender cannot directly observe the red agent's actions; instead it perceives only partial state transitions and indirect outcomes of attacks, making it difficult to reconstruct the true dynamics of the system. This partial observability poses challenges in predicting attacker's (red) actions, and thereby, inferring attacker's (red agent's) policy. These challenges significantly limit the defender from designing effective autonomous defense strategies. To overcome this limitation, we propose solutions to learn red agent policies and predict red actions at runtime.

A few notable works in the literature leveraged imitation learning to learn latent policies from observations in classical control environments~\cite{NEURIPS2022_efb2072a,pmlr-v97-edwards19a,NEURIPS2024} and in internet video~\cite{lapo}. However, none of these works are directly applicable to autonomous network environments. This is because the states in autonomous networks are typically discrete in nature and contain a snapshot of different ongoing network activity on each host/server.
Moreover, the state transitions in autonomous networks are dependent on both red agent (attacker's) actions and blue agent (defender's) actions in each timestep. Thus, we propose a \textit{Policy Learning Technique} 
that learns red agent policy and predicts red actions at runtime from network observations and blue  (defender) actions. 
Predicting red actions at runtime enables early detection of adversarial movements and estimation of attacker's intrusion levels over time before deep network compromise.

Reinforcement learning (RL) is widely used for autonomous cyber-defense, however, RL-based policies face challenges in scalability, interpretability, and adaptability in complex networks. 
Neurosymbolic methods address these limitations by combining learning with structured, interpretable reasoning. Recent works proposed behavior trees (BTs)~\cite{Li2021MixedDR,Colledanchise_2018} augmented with learning-enabled components (LECs), known as Evolving Behavior Trees (EBTs), for autonomous cyber defense~\cite{potteiger2024designing,oodd2025}. EBTs are modular and hierarchical, enabling them to encode explicit subtasks, capture control flows, and incorporate adaptive LECs for decision-making under uncertainty.
This design not only ensures the agent's ability to respond to diverse and dynamic attack patterns, but also ensures that the resulting models generalize effectively across simulation and real-world deployment. In this work, we extend the EBT proposed in \cite{oodd2025} by incorporating new behaviors to learn red agent policies and predict red actions at runtime.
To train and evaluate our approach, we use the CybORG CAGE Challenge 2 environment~\cite{cyborg_acd_2021,beyond-cage}, which leverages tactics and techniques from MITRE ATT\&CK~\cite{MITREATTACK} to create realistic cybersecurity environments for training red and blue agents.

Thus, the main contributions of our work are as follows.

\begin{enumerate}
\item We propose a \textit{Policy Learning Technique} that learns red agent (attacker) policies and predicts red (attacker) actions from observations and blue (defender) actions in a partially observable RL-based autonomous cyber-defense environment with discrete states and discrete actions. 

\item We integrate our proposed technique with a neurosymbolic behavior tree-based autonomous cyber-defense agent and demonstrate its ability to predict red actions in autonomous networks at runtime. 

\item We evaluate our proposed technique using CybORG CAGE Challenge 2~\cite{cyborg_acd_2021,beyond-cage}, a realistic cyber network simulation environment with tactics and techniques from MITRE ATT\&CK~\cite{MITREATTACK}. 
Our approach demonstrates high accuracy in predicting the red actions under different adversarial strategies and under dynamic adversarial strategy switching.
\end{enumerate}

\section{Autonomous Cyber-Defense Environment}\label{sec:background}
\noindent
The MITRE ATT\&CK knowledge base captures real-world adversarial techniques and provides a structured model of attacker behavior across domains such as Enterprise, Mobile, and Industrial Control Systems~\cite{MITREATTACK}. It systematically organizes adversarial activity into: Tactics, Techniques, Sub-techniques and Procedures.


The CybORG CAGE Challenge 2 interface, based on the MITRE ATT\&CK for Enterprise networks~\cite{kiely2023autonomous}, is designed to evaluate autonomous cyber-defense agents. The network consists of three subnets: Subnet~1 with five non-critical user hosts, Subnet~2 with three enterprise servers controlling the hosts in Subnet~1 and a host defending the network, and, Subnet~3 with three operational hosts and a critical operational server ensuring proper functioning of the network.

Each scenario run or episode in CybORG runs for fixed timesteps where red (attacker) and blue (defender) agents select actions from their respective action spaces. Each red action in CybORG is derived from the MITRE ATT\&CK knowledge base~\cite{cyborg_acd_2021}. A description of the red actions is given in Table~\ref{tab:redaction}. 
The red agent begins with an access to a user host and progresses through reconnaissance, exploitation, and privilege escalation to reach the enterprise servers and then the operational server, aiming to disrupt services via impact action~\cite{cyborg_acd_2021}.
\begin{table}[h]
    \centering
    \caption{Description of the red actions~\cite{MITREATTACK}.}
    \scalebox{0.95}{
    \begin{tabular}{|l|l|}\hline
        \textbf{Action} & \textbf{Purpose} \\ \hline
        Discover Remote  & ATT\&CK Technique T1018; Discovers new hosts/IP \\
         Systems & addresses in the network through active scanning.\\ \hline
        Discover Network & ATT\&CK Technique T1046; Discovers responsive \\
         Services&services on a selected host by initiating a connection\\
         & with that host.\\ \hline
        Exploit Network & ATT\&CK Technique T1210; Attempts to exploit a \\
        & specified service on a remote system.\\ \hline
        Escalate& ATT\&CK Tactic TA0004; Escalates the agent's\\
        &  privilege on the host\\ \hline
        Impact&ATT\&CK Technique T1489; Disrupts the performance  \\
        & of the network and fulfils the attacker's objective\\
        & of denying the operational service.\\ \hline
    \end{tabular}}
    \label{tab:redaction}
\end{table}

The blue agent mitigates threats using actions such as monitor, analyse, deploy decoys, remove threats, and restore~\cite{kiely2023autonomous}. The environment includes two types of attacker strategies: $Meander$ that explores and exploits the network one subnet at a time, seeking and gaining privileged access on all the hosts in a subnet before moving on to the next one, eventually reaching the Operational Server, and $B\_line$ that attempts to reach the Operational Server directly using prior knowledge of the network layout. Potteiger \emph{et al.} introduced a third red agent strategy, $RedSwitch$, that instantiates a red agent using $Meander$ and switches to $B\_line$ after a random number of timesteps~\cite{potteiger2024designing}.
\begin{figure*}
    \centering
    \includegraphics[width=\linewidth]{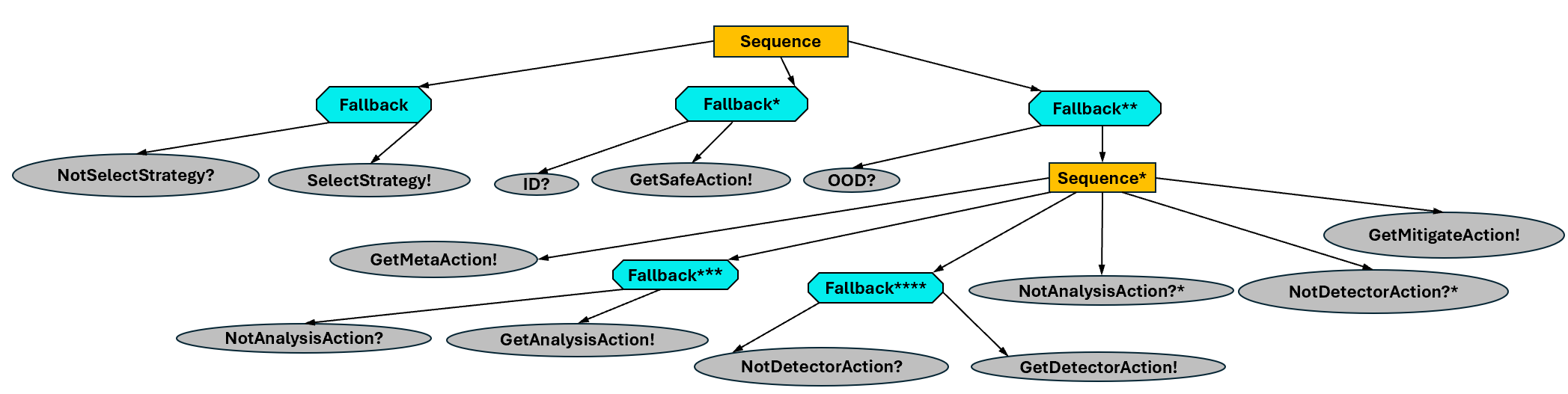}
    \caption{Evolving Behavior Tree (EBT) for Autonomous Cyber-Defense with online safety monitoring~\cite{oodd2025}.}
    \label{fig:ebt}
\end{figure*}

\section{Autonomous Cyber-Defense Agent using Evolving Behavior Trees}\label{sec:ebt}
\noindent
Neurosymbolic AI has emerged as a promising paradigm for autonomous cyber-defense, combining learning with structured reasoning to handle sophisticated attacks. Behavior Trees (BTs) serve as interpretable control structures that select defensive actions, and when augmented with learning-enabled components (LECs), they enable adaptive, high-level decision-making in dynamic environments. Their modularity supports easy integration of new behaviors for evolving threats. Prior work introduced Evolving Behavior Trees (EBTs), integrating BTs with LECs for adaptive defense~\cite{potteiger2024designing}, and extended them for runtime monitoring to detect anomalies and out-of-distribution states in cyber-defense systems~\cite{oodd2025}.

A Behavior Tree (BT) executes in discrete timesteps called ticks. At each tick, traversal starts from the root in a depth-first manner, with each node returning one of three statuses: $Running$ (execution ongoing), $Success$ (goal achieved), or $Failure$ (goal not met). Nodes are of two types: $Control$ behaviors and $Execution$ behaviors. $Control$ behaviors define execution flow: $Sequence$ (succeeds if all children succeed), $Fallback$ (succeeds if any child succeeds), and $Parallel$ (executes children simultaneously with configurable success conditions). $Execution$ behaviors are leaf nodes representing \emph{Conditions} (checks) or \emph{Actions} (tasks). During each tick, leaf node statuses propagate upward recursively, determining the overall outcome at the root.

Fig.~\ref{fig:ebt} presents the EBT from~\cite{oodd2025}, 
comprising six $Action$ behaviors: $SelectStrategy!$ (chooses defense strategy), $GetSafeAction!$ (restores system to a safe state), $GetMetaAction!$ (selects a defensive behavior), $GetAnalysisAction!$ (monitors and analyzes the network), $GetDetectorAction!$ (deploys detectors), and $GetMitigateAction!$ (removes threats or restores hosts). The EBT also includes five $Condition$ behaviors to ensure correct strategy selection, maintain operation within the training distribution, and validate execution of defensive actions. These behaviors map directly to the CybORG CAGE Challenge~2 environment: strategy selection aligns with controller policies, $GetSafeAction!$ maps to Restore, $GetAnalysisAction!$ to Monitor/Analyze, $GetDetectorAction!$ to DeployDecoy, and $GetMitigateAction!$ to Remove/Restore, enabling effective mitigation of adversarial activity.

\section{Learning from Observations}\label{sec:learning}
\noindent
Autonomous networks are partially observable,
as exemplified by CybORG CAGE Challenge 2 environment, which models real-world scenarios using the MITRE ATT\&CK. In such environment, the defender has no direct knowledge of attacker policies or actions and must infer system states by monitoring and analyzing hosts and network activity at each timestep.

\subsection{Problem Formulation}\label{subsec:prob}
\noindent
Our network can be represented as $\mathcal{N} = \{\mathcal{U} \cup \mathcal{E} \cup \mathcal{O}\}$, where $\mathcal{U} = \{u_1,u_2,\ldots,u_{n1}\}$ denotes the user subnet with $n1$ users, $\mathcal{E} = \{e_1,e_2,\ldots,e_{n2}\}$ denotes the enterprise subnet with $n2$ enterprise servers and hosts, and $\mathcal{O} = \{o_1,o_2,\ldots,o_{n3}\}$ denotes the operational subnet with $n3$ operational servers and hosts. There is a red agent (attacker) in the network with a set of $k$ possible red actions $RA = \{r_1,r_2,\ldots,r_k\}$ derived from tactics and techniques of MITRE ATT\&CK~\cite{MITREATTACK,cyborg_acd_2021,beyond-cage}. To defend the network, there is a blue agent (defender) residing in the network with a possible set of $m$ blue actions $BA = \{b_1,b_2,\ldots,b_m\}$~\cite{cyborg_acd_2021,beyond-cage}. Depending on the target on which red and blue actions are executed, we define red and blue action spaces as follows.

\begin{definition}
    Red Action Space: The red agent is associated with a red action space $A_r$ that is a combination of red action, $r_i \in RA$, and the host or subnet, $x$, on which the action is executed. $A_r: r_i \times x$, where $r_i \in RA$ and $(x = \mathcal{U})~\lor~(x = \mathcal{E})~\lor~(x = \mathcal{O})~\lor~(x \in \mathcal{U})~\lor~(x \in \mathcal{E})~\lor~(x \in \mathcal{O})$.
\end{definition}

\begin{definition}
    Blue Action Space: To mitigate red actions, the blue agent is associated with a blue action space $A_b$ that is a combination of blue action, $b_i \in BA$, and the host or server, $x$, on which the action is executed. $A_b: b_i \times x$, where $b_i \in BA$ and $(x \in \mathcal{U})~\lor~(x \in \mathcal{E})~\lor~(x \in \mathcal{O})$.
\end{definition}

Our network can be represented as a discrete-time Partially Observable Markov Decision Process (POMDP), \mbox{$\mathcal{M} := (S, A, O, T, R)$~\cite{mdp}}. Here, $S$ denotes the finite set of states, $A = \{A_r, A_b\}$ denotes the set of joint action spaces of the red and blue agents.
At timestep $t-1$, the red and blue agents perform the joint action $\vec{a}_{t-1} = \langle ra_{t-1}, ba_{t-1} \rangle$ from $A_r$ and $A_b$, respectively, resulting in the network state to transition from $s_{t-1}$ to $s_t$. $O$ denotes the finite set of observations for the red and blue agents. At timestep $t-1$, the joint observation $\vec{o}_{t-1} = \langle o^{r}_{t-1}, o^{b}_{t-1} \rangle$ denotes the observations available to the red and blue agents, respectively. $T$ denotes the transition probabilities between states, where $T(s', \vec{o}~|~s, \vec{a})$ denotes the probability to transition to state $s'$ producing the joint observation $\vec{o}$ given the current state $s$ and joint action $\vec{a}$. $R = \{R^r, R^b\}$ is the reward function for the red and blue agents, respectively. Note that our network is partially observable, \emph{i.e.}, the red observations and red actions are not observable to the blue agent. The main objective of the blue agent is to select a blue action at each timestep such that the cumulative blue rewards $\sum_{t=1}^{t\rightarrow\infty} R^b_{t-1}$ maximize over time. 

Although the RL-based defender (blue agent) executes actions at each timestep by scanning and analyzing the network states, it lacks knowledge of the red agent's policy or red actions. Therefore, our primary objective is to infer the red agent's policy and predict red actions using the blue observations and the blue action at each timestep.

\vspace{0.5em}
\noindent
\emph{Problem Statement: Given a network $\mathcal{N}$ with user subnet, enterprise subnet and operational subnet, a neurosymbolic autonomous cyber-defense agent (blue agent) trained with a RL policy $\pi$ against a given cyber-attacker (red agent), our objective is to learn the red agent's policy $\pi'$ and predict red action $ra_{t-1}$ at any timestep $t-1$ from blue observations in two consecutive timesteps, $o^b_{t-1}$ and $o^b_t$, and blue action $ba_{t-1}$ executed at timestep $t-1$.}

\begin{figure}
    \centering
    \includegraphics[width=\linewidth]{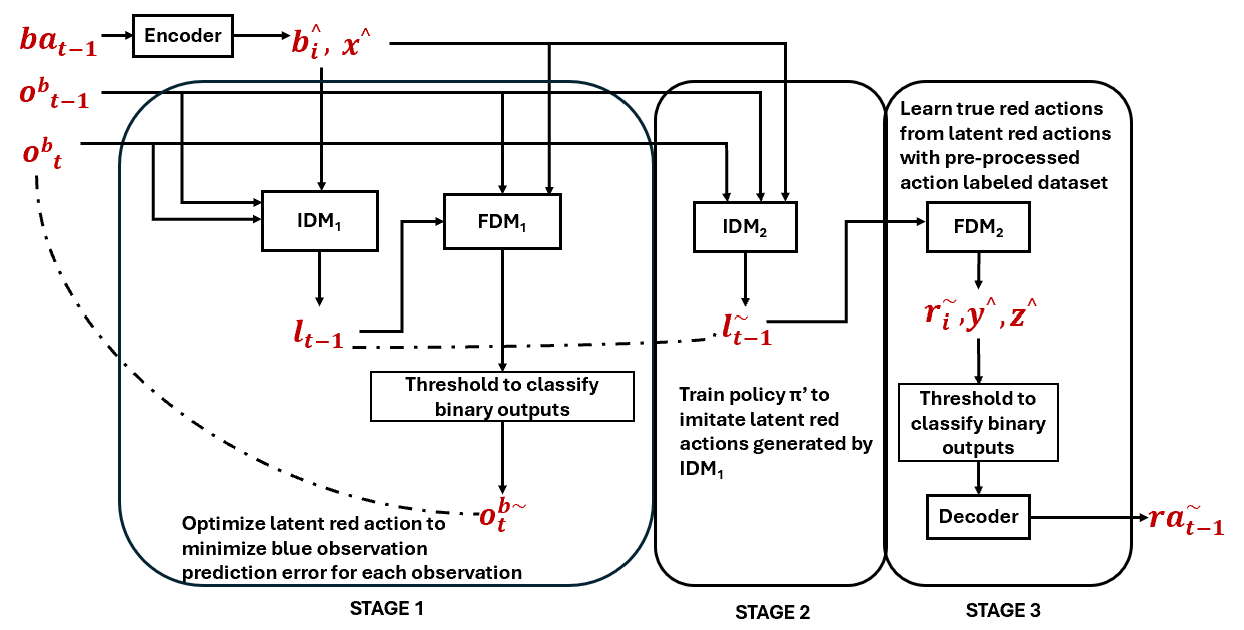}
    \caption{Three stages of learning Red Agent Policy $\pi'$ from blue observations and blue action.}
    \label{fig:policy}
\end{figure}

\subsection{Learning Red Agent Policy}\label{subsec:redaction}
\noindent
We propose a \textit{Policy Learning Technique} to learn a red agent policy $\pi'$ given blue agent policy $\pi$. Our proposed technique, at each timestep $t$, takes the previous and current blue observations ($o^b_{t-1}$ and $o^b_t$), and the previous blue action ($ba_{t-1}$) at timestep $t-1$ as inputs, and predicts the red action ($ra_{t-1}$) executed at $t-1$. Inferring red agent's policies under partial observability is challenging due to complex network dynamics. While prior work such as LAPO learns latent actions from video using imitation learning~\cite{lapo}, it is not directly applicable to our network as the blue observations in our network are discrete binary vectors, and each state transition depends on both observable blue action and non-observable red action.


Algorithm~\ref{alg:learn} highlights the key steps of our proposed \textit{Policy Learning Technique} which consists of three phases:

\noindent
\textbf{1. Data Generation Phase}: For a given blue agent policy $\pi$, we collect data, \emph{i.e.}, the blue observation ($o^b_t$), blue action ($ba_t$), and the blue observation transitions, $o^b_{t-1}, ba_{t-1} \rightarrow o^b_{t}$, at each timestep by interacting with our system to generate training dataset $\mathcal{D}$ for $\mathcal{T}$ timesteps, ($\mathcal{T}$ is a large number), over $N$ episodes (line~4 of Algorithm~\ref{alg:learn}). Additionally, we generate a red action labeled dataset $\mathcal{D}_R$ from the system for transitions in $\mathcal{D}$ (line~5 of Algorithm~\ref{alg:learn}). Note that these red action labels are not observable to the blue agent in autonomous networks as well as our system. We tweak our system to generate these labels only for training purpose. We generate dataset $\hat{\mathcal{D}_R}$ by pre-processing the action labels in $\mathcal{D}_R$ and splitting each red action label $ra_{t-1}$ into red action name, $r_i \in RA$, host name, $y \in \mathcal{N}$, and subnet $z$, where $z$ can be either $\mathcal{U}$, $\mathcal{E}$, or $\mathcal{O}$.

\begin{algorithm}
    \nonl{\textbf{Data Generation Phase}}\\
    \nl Assign $\mathcal{D},\mathcal{D}_R$,$\hat{\mathcal{D}_R}$ to $\{\}$\;
    \For{$e = 1, 2,\ldots, N$ episodes}
    {
        \For{$t = 1, 2, \ldots, \mathcal{T}$ timesteps}
        {
            Collect transitions $(o^b_{t-1},ba_{t-1} \rightarrow o^b_t)$ from the system for blue agent policy $\pi$ to generate $\mathcal{D}$\;
            Label $ra_{t-1}$ from the system for transitions in $\mathcal{D}$ to generate $\mathcal{D}_R$\;
            Pre-process $ra_{t-1} \in \mathcal{D}_R$ to generate $\hat{\mathcal{D}_R}$\;
        }
    }
    \nonl{\textbf{Policy Learning Phase}}\\
    \tcp{Encoding}
    \nl \For{each transition $(o^b_{t-1},ba_{t-1} \rightarrow o^b_t) \in \mathcal{D}$}
    {
        Split $ba_{t-1}$ into $b_i$ and $x \mid b_i \in BA$, $x \in \mathcal{N}$\;
        $Encode(b_i) \rightarrow \hat{b_i}$, $Encode(x) \rightarrow \hat{x}$\;
    }
    \tcp{Learn latent red actions}
    Model $IDM_1$ : $P_{IDM_1}(l_{t-1} \mid o^b_{t-1},o^b_t,\hat{b_i},\hat{x})$\;
    Model $FDM_1$ : $P_{FDM_1}(\widetilde{o^b_t} \mid o^b_{t-1},\hat{b_i},\hat{x},l_{t-1})$\; 
    Train $IDM_1$ and $FDM_1$ jointly over $\mathcal{D}$ to learn $l_{t-1}$\;
    \tcp{Train red agent policy $\pi'$}
    Model $IDM_2$ : $P_{IDM}(\widetilde{l_{t-1}} \mid o^b_{t-1},o^b_t,\hat{b_i},\hat{x})$\;
    Train $IDM_2$ over $\mathcal{D}$ to learn policy $\pi'$\;
    \tcp{Map latent red actions to red actions}
    Model $FDM_2$ : $P_{FDM_2}(\widetilde{r_i},\widetilde{y},\widetilde{z} \mid \widetilde{l_{t-1}})$\;
    Train $FDM_2$ over $\hat{\mathcal{D}_R}$\;
    Decode $(\widetilde{r_i},\widetilde{y},\widetilde{z}) \rightarrow \widetilde{ra_{t-1}}$\;
    \nonl{\textbf{Action Prediction Phase}}\\
    \nl Select $\pi'_i$ from $\{\pi'_1,\pi'_2,\ldots,\pi'_n\}$ corresponding to blue agent policy $\pi_i$\;
    $\pi'_i(o^b_{t-1},o^b_t,ba_{t-1}) \rightarrow \widetilde{ra_{t-1}} $\;
    return $\widetilde{ra_{t-1}}$\;       
\caption{\textit{Policy Learning Technique ($\pi$)}}
    \label{alg:learn}
\end{algorithm}

\noindent
\textbf{2. Policy Learning Phase}: Learning any red agent policy $\pi'$ involves the following steps as discussed below (see Fig.~\ref{fig:policy}).

\noindent
\textbf{I. Encoding}: For each transition $(o^b_{t-1},ba_{t-1} \rightarrow o^b_t)$ in $\mathcal{D}$, the blue action ($ba_{t-1}$) is decomposed into its action type $b_i \in BA$ and target host $x \in \mathcal{N}$. These components are then one-hot encoded w.r.t. the blue action set $BA$ and the set of hosts $\mathcal{N}$, producing encoded blue action $\hat{b_i}$ and encoded host $\hat{x}$.

\noindent
\textbf{II. Learning latent red actions}: This stage jointly trains two dynamics models to learn latent red actions ($l_{t-1}$)~(lines 10-12 of Algorithm~\ref{alg:learn}). The first is an inverse-dynamics model ($IDM_1$) which estimates $P_{IDM_1}(l_{t-1} \mid o^b_{t-1},o^b_{t},\hat{b_i},\hat{x})$. The second is a forward-dynamics model ($FDM_1$) which predicts the current blue observation as $P_{FDM_1}(\widetilde{o^b_t} \mid o^b_{t-1},\hat{b_i},\hat{x},l_{t-1})$. 
    
For each transition $(o^b_{t-1},ba_{t-1} \rightarrow o^b_t)$ in $\mathcal{D}$, the $IDM_1$ takes $o^b_{t-1}$, $o^b_t$, and $(\hat{b_i}, \hat{x})$ as input to infer the latent red action $l_{t-1}$. The $FDM_1$ then uses $o^b_{t-1}$, $(\hat{b_i}, \hat{x})$ and $l_{t-1}$ to predict the current observation $\tilde{o^b_t}$. Both the models are trained jointly via stochastic gradient descent to minimize the observation error $|\tilde{o^b_t} - o^b_t|$, thereby learning meaningful latent red actions. Since the blue observations in our networks are discrete binary vectors, we set a threshold to categorize the predicted observation vectors in binary.

\noindent
\textbf{III. Training to learn Red Agent Policy \boldmath{$\pi'$}}: This stage trains a second inverse dynamics model $IDM_2$ following $P_{IDM_2}(\widetilde{l_{t-1}} \mid o^b_{t-1},o^b_{t},\hat{b_i},\hat{x})$ to imitate the latent red actions learned by $IDM_1$ (lines 13–14 of Algorithm~\ref{alg:learn}). 
Trained via stochastic gradient descent, $IDM_2$ learns policy $\pi'$ by imitating latent red actions, minimizing the loss $|\widetilde{l_{t-1}} - l_{t-1}|$.

\noindent
\textbf{IV. Mapping latent red actions to actual red actions in \boldmath{$A_r$}}: The predicted latent red actions are continuous vectors. To map them to  discrete red actions in  $A_r$, we train a forward dynamics model $FDM_2$ defined as $P_{FDM_2}(\widetilde{r_i},\widetilde{y},\widetilde{z} \mid \widetilde{l_{t-1}})$ using the action-labeled preprocessed dataset $\hat{\mathcal{D}_R}$~(lines 15-16 of Algorithm~\ref{alg:learn}). $FDM_2$ learns to map each latent red action to its corresponding discrete action components. The outputs of $FDM_2$ are continuous vectors, which are thresholded to obtain binary one-hot encodings. Each predicted red action comprises three components: the action name $\widetilde{r_i}$ from $RA$, the target host $\widetilde{y}$ and the subnet $\widetilde{z}$. A decoder is used to segment and map these outputs into the final predicted red action, $\widetilde{ra_{t-1}} \in A_r$~(line 17 of Algorithm~\ref{alg:learn}).

Note that, all dynamics models across the training stages are implemented as deep neural networks (DNNs), with different model specifications for each stage.

\noindent
\textbf{3. Prediction Phase}: Given a set of $n$ trained blue agent policies, $\{\pi_1,\pi_2,\ldots,\pi_n\}$, we learn a corresponding set of red agent policies, $\{\pi'_1,\pi'_2,\ldots,\pi'_n\}$, offline using the proposed framework. At runtime, the system selects the appropriate red policy $\pi'_i$ based on the active blue policy $\pi_i$. At each timestep $t$, the selected policy takes previous blue observation $o^b_{t-1}$, current blue observation $o^b_t$, and previous blue action $ba_{t-1}$ as input to predict the red action executed at timestep $t-1$.

\begin{figure*}
    \centering
    \includegraphics[width=\linewidth]{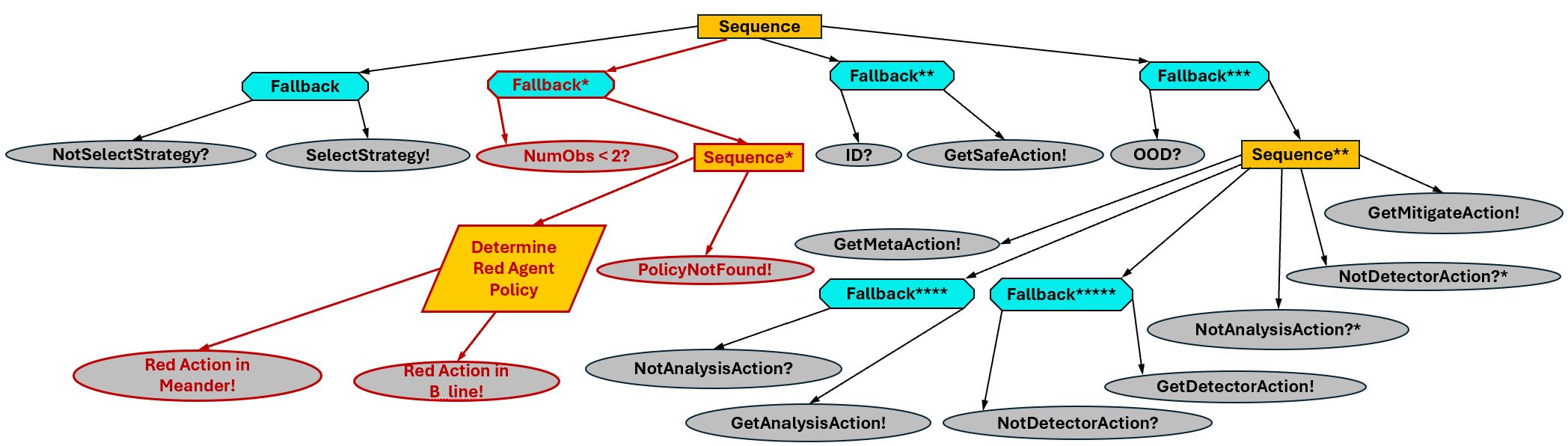}
    \caption{EBT in \cite{oodd2025} with  newly added behaviors (nodes marked in red) to predict red actions at runtime.}
    \label{fig:ebt_update}
\end{figure*}

\begin{table*}
    \centering
    \caption{Model specifications of the three stage dynamics models for training different red strategies.}
    \scalebox{0.96}{
    \begin{tabular}{|l|l|l|l|l|}\hline
    \textbf{Stage}&\textbf{Model}&\textbf{Input}&\textbf{Architecture}&\textbf{Output}\\ \hline  
    Stage~1 (Learn& $IDM_1$ &$o^b_{t-1},o^b_{t},ba_{t-1}$&DNN with $3$ hidden layers each of size $150$, and ReLU activation in each layer&$l_{t-1}$ (intermediate)\\ \cline{2-5}    
    (latent red actions)&\multirow{2}{*}{$FDM_1$}&\multirow{2}{*}{$o^b_{t-1},ba_{t-1},l_{t-1}$}&DNN with $3$ hidden layers each of size $120$, ReLU activation in each layer and &\multirow{2}{*}{$\tilde{o^b_t}$}\\
    &&& Softplus in the last layer&\\ \hline
    Stage~2 (Train &\multirow{2}{*}{$IDM_2$}&\multirow{2}{*}{$o^b_{t-1},o^b_t,ba_{t-1}$}&DNN with $5$ hidden layers each of size $150$, and ReLU activation in each layer&\multirow{2}{*}{$\widetilde{l_{t-1}}$}\\
    red policy)&&& &\\ \hline
    Stage~3 (Predict&\multirow{2}{*}{$FDM_2$}&\multirow{2}{*}{$\widetilde{l_{t-1}}$}&DNN with $6$ hidden layers each of size $15$, ReLU activation function in each &\multirow{2}{*}{$\widetilde{ra_{t-1}}$}\\
     red actions)&&& layer and Softplus in the last layer& \\ \hline 
    \end{tabular}}
    \label{tab:specifications}
\end{table*}

\section{Incorporation of Red Action Prediction Behavior in the Evolving Behavior Tree}\label{sec:integration}
\noindent
We extend the EBT proposed in \cite{oodd2025} by incorporating red action prediction behaviors and learning-enabled components (LECs) to enable red action prediction at every tick for different red agent policies. As shown in Fig.~\ref{fig:ebt_update}, we introduce one Condition and three Action behaviors (marked in red). The condition $NumObs < 2?$ ensures availability of at least one prior observation ($o^b_{t-1}$) before prediction. Leveraging the $SelectStrategy!$ behavior \cite{potteiger2024designing}, the defender identifies the active blue policy based on adversarial behavior and accordingly selects the learned red policy. The Action behaviors, $Red~Action~in~Meander!$ and $Red~Action~in~B\_line!$, predict red actions under $B\_line$ and $Meander$ respectively. If no learned red policy exists, $PolicyNotFound!$ is triggered to log state transitions for offline analysis.

\begin{table}
    \centering
    \caption{Prediction Accuracy of our models under different red strategies.}
    \begin{tabular}{|l|l|l|l|}\hline
        \textbf{Red Policy} & \textbf{Stage~1} & \textbf{Stage~2}& \textbf{Stage~3}\\ \hline
         \multirow{2}{*}{\textcolor{red}{\boldmath{$B\_line$}}}&$\mathbf{99.93\%}$&All latent&$\mathbf{95.28\%}$ (host)\\
         & &red actions&$\mathbf{99.82\%}$ (subnet) \\ \cline{1-2}\cline{4-4}
         \multirow{2}{*}{\textcolor{red}{\boldmath{$Meander$}}}&$\mathbf{94.96\%}$&are learnt&$\mathbf{77.71\%}$ (host) \\
         & &with an & $\mathbf{97.49\%}$ (subnet) \\ \cline{1-2}\cline{4-4}
         \multirow{2}{*}{\textcolor{red}{\boldmath{$RedSwitch$}}}&$\mathbf{95.03\%}$&error bound&$\mathbf{83.83\%}$ (host) \\
         & &of $\mathbf{0.1}$&$\mathbf{95.85\%}$ (subnet) \\ \hline    
    \end{tabular}
    \label{tab:accuracy}
\end{table}

\begin{table}
    \centering
    \caption{Blue Observation Prediction Error under Different Red Strategies.}
    \begin{tabular}{|c|c|c|c|c|c|}\hline
    \multirow{2}{*}{\textbf{Red Policy}}&\multicolumn{5}{|c|}{\textbf{Blue Observation Prediction Error}}\\ \cline{2-6}
    &\textbf{1-bit}&\textbf{2-bit}&\textbf{3-bit}&\textbf{4-bit}&\textbf{5-bit}\\ \hline
    \textcolor{red}{\boldmath{$B\_line$}}& \boldmath{$0.02\%$}&\boldmath{$0.04\%$}&\boldmath{$0\%$}&\boldmath{$0\%$}&\boldmath{$0\%$}  \\ \hline
    \textcolor{red}{\boldmath{$Meander$}} & \boldmath{$4.48\%$}&\boldmath{$0.27\%$}&\boldmath{$0.11\%$}&\boldmath{$0.07\%$}&\boldmath{$0.07\%$} \\ \hline
    \textcolor{red}{\boldmath{$RedSwitch$}}&\boldmath{$2.28\%$}&\boldmath{$1.51\%$}&\boldmath{$0.64\%$}&\boldmath{$0.49\%$}&\boldmath{$0.03\%$}\\ \hline
    \end{tabular}
    \label{tab:prederr}
\end{table}

\begin{table*}
    \centering
    \caption{Predicted Red Actions with target host/subnet and frequencies under Privilege-Level Compromise for Different Red Strategies.}
    \begin{tabular}{|l|l|l|l|l|}\hline
    \multirow{3}{*}{\textbf{Red Policy}}&\multicolumn{4}{|c|}{\textbf{Predicted Red Actions with target host/subnet and their frequencies}}\\ \cline{2-5}
    & \textbf{DiscoverNetworkServices} & \textbf{DiscoverRemoteSystems}&\textbf{ExploitRemoteService}&\textbf{PrivilegeEscalate}\\ \hline
    \multirow{2}{*}{\textcolor{red}{$B\_line$}}&Enterprise0: 203, Enterprise2: 16, &Enterprise0: 16, &User3: 190, Enterprise0: 257, &User3: 180, User2: 1, \\ 
     &Op\_Server0: 3 &Enterprise1: 2 &Enterprise2: 16  & Enterprise0: 16, \\ 
     & & & & Enterprise1: 3, \\
     & & & & Enterprise2: 3 \\ \hline
    \multirow{2}{*}{\textcolor{red}{$Meander$}}& Enterprise1: 7& &User2: 2, User1: 1, User4: 1, & User3: 47, User4: 1,\\ 
    & & & User3: 1, Enterprise0: 1&Enterprise1: 21 \\ \hline
    \multirow{2}{*}{\textcolor{red}{$RedSwitch$}}& User3: 3, Enterprise0: 143, &Enterprise0: 18, &User3: 132, Enterprise0: 187, & User3: 171, User4: 2, User2: 2, \\
    & Enterprise1: 5, Enterprise2: 17, &Enterprise1: 1 &Enterprise2: 16, Enterprise1: 1 & Enterprise0: 18, Enterprise2: 4 \\ 
    & Op\_Server0: 4 & & & \\ \hline     
    \end{tabular}
    \label{tab:actionanalysis}
\end{table*}

\section{Experiments and Evaluation}\label{sec:experiments}
\noindent
We evaluate our proposed \textit{Policy Learning Technique} in the CybORG CAGE Challenge 2 environment. To enable runtime prediction of red actions, we extend the EBT from \cite{oodd2025} by integrating learned red agent policies and red action prediction behaviors. We evaluate our approach across two distinct red agent strategies and under dynamic red strategy switching.

\subsection{Experimental Setup}\label{subsec:setup}
\noindent
For evaluation, we develop a software architecture extended from~\cite{oodd2025} that utilizes PyTrees and executes the Blue EBT agent, Runtime Monitoring, and the learned red policies on CybORG CAGE Challenge 2. Communications between the simulator and the EBT occurs via a blackboard interface to prevent data leakages. We perform experiments on a Linux machine ($2.1$ GHz Intel Xeon, $32$ GB RAM). We train our models to learn two red agent policies by generating dataset $\mathcal{D}$, consisting of $10,000$ transitions per strategy ($100$ episodes $\times$ $100$ timesteps). From $\mathcal{D}$, we derive the action-labeled dataset $\mathcal{D}_R$ and processed it to generate $\hat{\mathcal{D}_R}$ for model training. 

We employ a three-stage dynamics model (Table~\ref{tab:specifications}) optimized via stochastic gradient descent, utilizing Cross Entropy for Stages~1 and~3, and Mean Absolute Error for Stage~2. Blue actions are decomposed into action name and host, and encoded using one-hot encoding. In Stage~1, the inverse dynamics model ($IDM_1$) takes inputs of size $130$, derived from two consecutive binary state vectors (size $52$ each), $13$ blue actions, and $13$ hosts. The forward model ($FDM_1$) input size varies with latent action dimensions, which are tuned between $10$-$50$; optimal sizes are $18$ ($B\_line$) and $29$ ($Meander$), reflecting higher variability in Meander behavior. Stage~2 uses $IDM_2$ with the same input size ($130$) to imitate latent actions. In Stage~3, $FDM_2$ maps latent actions to discrete outputs, with input sizes $18$ and $29$ for $B\_line$ and $Meander$, respectively.

Training uses mini-batches of size $10$ (with replacement) for $1000$ epochs at a learning rate of $0.01$. Outputs are thresholded to obtain binary representations. The final output is a $21$-bit vector: $5$ bits for action type, $3$ for subnet, and $13$ for host. A decoder maps these into executable red actions. These trained models are integrated as LECs within the EBT, enabling runtime prediction of red actions using consecutive blue observations and encoded blue actions. The link to our codebase is available at \cite{code}.


\subsection{Evaluation and Results}
\noindent
We evaluate our models under two adversarial strategies, $B\_line$ and $Meander$, and further assess generalization under the $RedSwitch$ strategy without additional training. For each policy, we conduct $10$ test runs, each test run comprising $10$ episodes of $100$ timesteps each. Performance is measured using \textit{Prediction Accuracy} (the ratio of correct predictions to total predictions) across predicted observations, predicted latent red actions, and predicted red actions. 

\subsubsection{Evaluation using Prediction Accuracy}
Table~\ref{tab:accuracy} reports the average prediction accuracy across the three stages over $10$ test runs. The highest accuracy for both blue observation and red action prediction is observed under $B\_line$ due to its more deterministic and goal-oriented behavior as compared to higher variability under $Meander$ and $RedSwitch$ strategies.

Table~\ref{tab:prederr} presents bit-level blue observation prediction errors (in percentage) in Stage~1 under $B\_line$, $Meander$ and $RedSwitch$ strategies. Errors are minimal under $B\_line$ (upto $2$ bit errors), while $Meander$ and $RedSwitch$ exhibit higher errors (upto $5$ bit errors). Although $Meander$ shows a higher percentage of $1$-bit prediction errors than $RedSwitch$, it shows less percentage of $n$ bit errors ($n > 1$). This is because, under red strategy switching, our network transitions to new states over some timesteps which are not in the training distribution. 

From Table~\ref{tab:accuracy}, it is evident that the latent red actions in Stage~2 are predicted with a low error bound ($\leq 0.1$) across all cases. In Stage~3, the red action prediction accuracy (at host and subnet levels) is highest for $B\_line$, reflecting its deterministic nature, while red action prediction is less for $Meander$ due to higher randomness. Under $RedSwitch$, the red action prediction accuracy (host level) remains low for timesteps during which the red agent follows $Meander$ strategy. However, on switching to $B\_line$, prediction accuracy increases which increases the overall prediction accuracy over an episode as compared to $Meander$. 
Also, across all three strategies, subnet-level red action prediction accuracy consistently exceeds host-level accuracy. This is due to the inherent randomness in red agents' selection of target hosts, which makes fine-grained (host-level) prediction more challenging.

\subsubsection{Evaluation using Intrusion Level Measure} The red actions predicted by our proposed technique can be further analyzed to assess the intrusion level of different red agents in the network. This analysis further enables: 1) Evaluation of blue agent's effectiveness and refinement of mitigation strategies if needed, and 2) Vulnerability analysis to identify weaknesses exposed by the red agent activity.

In CybORG CAGE Challenge 2, the blue observations are represented as $52$-bit binary vectors, with four bits per host/server. Two of these bits encode compromise levels: `00' (No Compromise), `01' (User-Level Compromise), `11' (Privilege-Level Compromise). We quantify intrusion by assigning an intrusion score $\mathcal{I}_i$ to each host/server $i$, ($i \in \mathcal{N}$).  

\[
\mathcal{I}_i =
\begin{cases} 
{\footnotesize 0} & \text{\footnotesize if $i$ is in No~Compromise} , \\
{\footnotesize 1} & \text{\footnotesize if $i$ is in User-Level~Compromise} , \\
{\footnotesize 10} & \text{\footnotesize if $i$ is in Privilege-Level~Compromise} 
\end{cases}
\]

The total network intrusion at each timestep is $\sum_{i=1}^{|\mathcal{N}|} \mathcal{I}_i$.

Table~\ref{tab:actionanalysis} shows predicted red actions, their target host/subnet and red action frequencies under Privilege-Level Compromise out of $100$ episodes each with $100$ timesteps. From 
Table~\ref{tab:actionanalysis}, our network undergoes Privilege-Level Compromise for $82$ timesteps under $Meander$,  $903$ timesteps under $B\_line$ and $724$ timesteps under $RedSwitch$. These results indicate that our blue EBT agent is significantly more effective in mitigating attacks under $Meander$ strategy as compared to $B\_line$ and $RedSwitch$. By analyzing the blue observations, predicted red actions and network parameters in these timesteps, we can identify the weaknesses in our network and refine our mitigation actions accordingly.

\section{Related Works}\label{sec:related}
\noindent
The cybersecurity landscape is rapidly evolving as attacks become more sophisticated. The MITRE ATT\&CK knowledge base provides a foundation for modeling adversarial behavior~\cite{MITREATTACK}, while the CybORG CAGE Challenge 2 serves as a widely used environment for training and evaluating RL-based defense agents~\cite{kiely2023autonomous}. Prior work has explored hierarchical RL~\cite{Foley2022}, ensemble methods~\cite{beyond-cage}, curriculum learning~\cite{2021arXiv210307583M}, and reward shaping~\cite{Reward2023Bates} to address sparse, penalty-driven feedback in cyber environments.

Parallel efforts in neurosymbolic AI combine neural learning with symbolic reasoning for robust and interpretable cyber defense~\cite{neurosymbolic2023}. Behavior Trees (BTs), in particular, enable modular and interpretable policy design, often integrated with RL or HRL~\cite{Li2021MixedDR}. Extensions such as Evolving Behavior Trees (EBTs)~\cite{potteiger2024designing} and runtime monitoring frameworks~\cite{oodd2025} improve adaptability and safety in partially observable settings. However, these approaches do not explicitly model or predict attacker (red agent) policies.

In autonomous network defense, partial observability remains a key challenge, as defenders must act with limited visibility while attacker actions remain hidden. Although imitation learning has been used to infer latent policies in other domains~\cite{lapo,pmlr-v97-edwards19a,NEURIPS2022_efb2072a,NEURIPS2024,mosbach2025learning}, none of these works directly apply to discrete, partially observable cyber environments. 

\section{Conclusion and Future Works}
\noindent
Autonomous networks increasingly rely on RL-based neurosymbolic agents for cyber defense, but partial observability limits a defender's ability to infer attacker behavior. To address this, we propose a \textit{Policy Learning Technique} that uses imitation learning to infer red agent policies and predict red actions at runtime from consecutive observations and blue actions. Beyond cyber defense, our proposed technique can also be adopted to learn a policy for any partially observable RL-based agent with discrete states and discrete actions.
We integrate this technique into a behavior tree-based framework, enabling proactive defense and improved intrusion assessment. Evaluations in the CybORG CAGE Challenge 2 environment across multiple adversarial strategies show high prediction accuracy and effective intrusion-level estimation. However, as this environment is simulated (based on the MITRE ATT\&CK model), real-world networks may exhibit more complex dynamics. Future work will focus on validating the approach in realistic emulated environments and leveraging predicted red actions to train more adaptive and robust blue agents.

\section{Acknowledgment}
\noindent
This material is based on research sponsored by DARPA under agreement number HR001124C0425. The U.S. Government is authorized to reproduce and distribute reprints for Governmental purposes notwithstanding any copyright notation thereon. The views and conclusions contained herein are those of the authors and should not be interpreted as necessarily representing the official policies or endorsements, either expressed or
implied, of DARPA or the U.S. Government.

\bibliographystyle{plain} 
\bibliography{ref}

\end{document}